\documentstyle[preprint,aps,floats,epsf]{revtex}

\newcommand{ \slashchar }[1]{\setbox0=\hbox{$#1$}   
   \dimen0=\wd0                                     
   \setbox1=\hbox{/} \dimen1=\wd1                   
   \ifdim\dimen0>\dimen1                            
      \rlap{\hbox to \dimen0{\hfil/\hfil}}          
      #1                                            
   \else                                            
      \rlap{\hbox to \dimen1{\hfil$#1$\hfil}}       
      /                                             
   \fi}                                             %

\newcommand{\cM}{{\cal M}}
\newcommand{\vn}{{\vec n}}
\newcommand{\bei}{{\beta_i}}
\newcommand{\bef}{{\beta_f}}
\newcommand{\gkk}{G_{KK}^{}}
\newcommand{\dep}{{\dot{\varepsilon}}}
\def\lsim{\mathrel{\raise.3ex\hbox{$<$\kern-.75em\lower1ex\hbox{$\sim$}}}}
\def\gsim{\mathrel{\raise.3ex\hbox{$>$\kern-.75em\lower1ex\hbox{$\sim$}}}}

\begin{document}

\tighten
\preprint{ \vbox{
\hbox{MADPH--99--1118}
\hbox{hep-ph/9905474}}}
\title{Astrophysical Constraints on Large Extra Dimensions}
\author{V. Barger, T. Han, C. Kao and R.-J. Zhang}
\address{Department of Physics, University of Wisconsin\\ 
1150 University Avenue, Madison, WI 53706, USA }
\date{May, 1999}

\maketitle

\begin{abstract}
In the Kaluza-Klein (KK) scenario with $n$ large
extra dimensions where
gravity propagates in the $4+n$ dimensional bulk of
spacetime while gauge and matter fields are confined
to a four dimensional subspace, the light graviton KK modes 
can be produced in the Sun, red giants 
and supernovae. We study the energy-loss rates through
photon-photon annihilation, electron-positron annihilation, 
gravi-Compton-Primakoff scattering, gravi-bremsstrahlung 
and nucleon-nucleon bremsstrahlung, and derive lower limits to the 
string scale $M_S$. The most stringent lower limit obtained
from SN1987A leads to $M_S> 30 - 130$ TeV ($2.1-9.3$ TeV)
for the case of two (three) large extra dimensions.
\end{abstract}

\vspace{2cm}

\vfill

\pagebreak

\section{Introduction}

Recently there has been revived interest in physics of extra
spatial dimensions.  Compact spatial dimensions with inverse radius
at the order of the grand unified scale $\sim 10^{16-17}$ GeV 
are familiar ingredients in string compactifications and 
have been studied extensively since the mid-80's \cite{string}. 
However, recent developments 
in string duality suggest that it is possible to have a much 
lower string or compactification scale \cite{dual}. 
In particular, it is conceivable to set the scale at the
order of a TeV, corresponding to a weak-scale string 
theory \cite{tevstring}. 
Such a low string scale has the phenomenological
attraction of a lighter and experimentally accessible string 
state spectrum. Furthermore if the large Planck mass 
is attributed to the existence of $n$
extra dimensions, then the sizes of these extra dimensions ($R$) 
can be in the range of 1 fm to 1 mm for $n=6$ to 2 \cite{dimo}. 
The case of one large extra dimension implies modifications 
of Newton's law in the range of earth-sun
distance and is therefore excluded. For $n=2$ with 
$R\sim {\cal O}(1\ {\rm mm})$, it might be probed with 
laboratory gravitational experiments \cite{gravi}.

We consider the scenario that only 
gravity propagates in the extra dimensions, while the Standard Model
(SM) fields and interactions are ``confined'' to a four-dimensional 
subspace. 
In this scenario, the effect of large extra dimensions
arises only from interactions involving
the Kaluza-Klein (KK) excitations of 
the gravitons from compactification. 
At an energy scale much lower than the string scale,
one can construct an effective theory of KK gravitons
interacting with the standard model fields \cite{hlz}. 
Each graviton KK state couples to the SM field with
the gravitational strength according to
\begin{equation}
{\cal L}\ = -{\kappa\over2}\sum_{\vec n}
\int d^4 x\ h^{\mu\nu, {\vec n}} T_{\mu\nu}\ ,
\label{inter}
\end{equation}
where $\kappa=\sqrt{16\pi G_N}$, and
the summation is over all KK states labeled by the level $\vec n$.
$T_{\mu\nu}$ is the energy-momentum tensor of the SM 
and $h^{\mu\nu, {\vec n}}$ the KK state with mass 
$$m_{\vec n}^2={\vec n}^2/R^2\ .$$
We note that Eq.~(\ref{inter}) is of the same form as that for
massless graviton-matter coupling. Since for large $R$ the KK
gravitons are very light, they may be copiously produced in 
high energy processes. For real emission of the KK gravitons 
from a SM field, the total cross-section can be written as
\begin{equation}
\sigma_{\rm tot}\ =\ \kappa^2\sum_{\vec n} \sigma({\vec n})\ ,
\end{equation}
where the dependence on the gravitational coupling is factored out.
Because the mass separation of adjacent KK states,  ${\cal O}(1/R)$,
is usually much smaller than typical energies in a physical process,
we can approximate the summation by an integration.
Identifying the relation between the Planck mass in 4-dimension
($M_{\rm Pl}$) and that in $(4+n)$-dimension ($M_S$) according to 
\begin{equation}
\Omega_n M^{-2}_{\rm Pl} R^n \ =\ M_S^{-(n+2)}\ ,
\label{rel}
\end{equation}
where $\Omega_n$ is the $n$-dimensional spherical volume,
one can immediately infer that $\sigma_{\rm tot}$ has an 
$M_S^{-(n+2)}$ dependence. Thus the large degeneracy of the KK states 
compensates for the weakness of a single KK interaction. 
The associated rich collider phenomenology has been
the topic of many recent studies \cite{hlz,nima,collider,phenostring}.

In this paper, we study astrophysical consequences of this 
scenario in the effects of KK graviton emission in hot stars such as
the Sun, red giants and supernova SN1987A. 
As in the classic example of an invisible axion \cite{axi}, 
excessive energy losses in the stars can 
alter the stellar evolution and severe 
constraints can thereby be placed on any weakly-interacting 
light particle like the KK graviton.
We first compute in Sec.~2 the energy-loss rate for various 
processes involving emission of the KK gravitons $(\gkk)$, 
which include
\begin{itemize}
\item[(a)] $\gamma\gamma \to \gkk$, Photon-photon annihilation;
\item[(b)] $e^-e^+ \to \gkk$, Electron-positron annihilation;
\item[(c)] $e^-\gamma \to e^-\ \gkk$, Gravi-Compton-Primakoff scattering;
\item[(d)] $e^- (Ze) \to e^- (Ze)\ \gkk$, Gravi-bremsstrahlung in a 
static electric field;
\item[(e)] $NN \to NN\ \gkk$, Nucleon-nucleon bremsstrahlung.
\end{itemize}
We then use these formulae to derive lower limits on the string scale 
$M_S$ in Sec.~3. We summarize our results and conclude in Sec.~4. 

Many of the processes listed above were considered first 
in Ref.~\cite{nima}, with rate estimates based only on dimensional 
analysis. When our calculation was in progress, another related
work appeared \cite{CP}, in which the nucleon-nucleon bremsstrahlung 
process was studied in detail. Our results for this process are 
consistent with their calculations.

\section{Star Energy-loss via KK gravitons}

Weakly-interacting light particles may result in energy losses 
for hot stellar objects such as the Sun, red giants and 
SN1987A, with the invisible axion as the classic
example \cite{axi,PR,review}. 
We here study the energy loss due to the escaping KK gravitons. 
An important difference from the axion case is that the KK graviton
and matter interactions are of gravitational strength,
so the KK states never become thermalized and always freely escape.
In this Section, we calculate the volume energy-loss rates (emissivities) 
for various processes via KK graviton emission. 
The energy-loss rates have a high-power dependence on the string
scale, namely of $M_S^{-(n+2)}$, and corrections to our approximate
calculations would not significantly alter the lower limits on $M_S$
that we obtain.

\subsection{Photon-photon annihilation to $KK$ gravitons}

\begin{figure}[thb]
\epsfysize=1.5in
\epsffile[80 360 400 460]{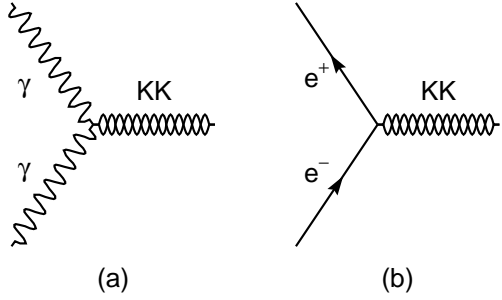}
\caption[]{Feynman diagrams for (a) photon-photon and 
(b) electron-positron annihilation into a KK graviton.
We represent KK gravitons by double-sinusoidal curves.
\label{pp}}
\end{figure}

Photons are abundant in hot stars.
We first consider photon-photon annihilation to a KK graviton.
Unlike the invisible axion, the KK gravitons couple to photons
at the tree-level, as shown in Fig.~\ref{pp}(a). 
Using the Feynman rules derived in Ref. \cite{hlz}, the spin-averaged
total cross-section for this process is easily found to be
\begin{equation}
\sigma_{\gamma\gamma\rightarrow \gkk}(s,m_{\vec n})\ =\ 
{\pi\kappa^2 \sqrt{s}\over 16}
\delta (m_{\vec n}-\sqrt{s})\ ,
\end{equation}
where $s$ is the center of mass energy, and $m_{\vec n}$ the mass
of the KK state at level $\vec n$.

The volume emissivity of a hot star with a temperature $T$ 
through this process is obtained by thermal-averaging over 
the Bose-Einstein distribution \footnote{This expression 
is similar to that of the energy loss rate 
via $\gamma\gamma \to \nu\bar{\nu}$ \cite{ggnn}.}
\begin{equation}
Q_\gamma\ =\ \int {2 d^3{\vec k}_1\over (2\pi)^3} {1\over e^{\omega_1/T}-1}
\int {2 d^3{\vec k}_2\over (2\pi)^3} {1\over e^{\omega_2/T}-1}
{s(\omega_1+\omega_2)\over 2\omega_1\omega_2}\sum_{\vec n}  
\sigma_{\gamma\gamma\rightarrow \gkk}(s,m_{\vec n}),
\end{equation}
where the summation is over all KK states, and the squared center 
of mass energy $s$ is related to the photon energies 
and the angle between the two photon momenta $\theta_{\gamma\gamma}$
as follows:
\begin{equation}
s\ =\ 2 \omega_1\omega_2 (1-\cos\theta_{\gamma\gamma})\ .
\end{equation}
After carrying out the integrals and the summation over KK states, we find
\begin{equation}
Q_\gamma\ =\ {2^{n+3}
\Gamma({n\over2}+3)\Gamma({n\over2}+4)\zeta({n\over2}+3)\zeta({n\over2}+4)
\over (n+4)\pi^2}
{T^{n+7}\over M_S^{n+2}} \ ,
\end{equation}
where we have used Eq. (\ref{rel}). Numerically, these 
Riemann zeta-functions are close to 1.
In this calculation, we have neglected the plasma effect, through
which the photons can have different energy 
dispersion relations from those of free particles \cite{PR}.

\subsection{Electron-positron annihilation to $KK$ gravitons}

In supernovae, the core temperature ($T_{\rm SN}$)
is high enough for pair-creation of electrons and positrons,
which subsequently annihilate to KK gravitons, as depicted  
in Fig.~\ref{pp}(b), with a total cross-section 
(neglecting the electron mass since $m_e\ll T_{\rm SN}$)
given by
\begin{equation}
\sigma_{e^-e^+\rightarrow \gkk}(s,m_{\vec n}) \ =\ 
{\pi\kappa^2\sqrt{s}\over 64} 
\delta (m_{\vec n}-\sqrt{s})\ .
\end{equation}

The volume emissivity is obtained by thermal-averaging over the
Fermi-Dirac distribution
\begin{eqnarray}
Q_e &=&  \int {2 d^3{\vec k}_1\over (2\pi)^3} {1\over e^{(E_1-\mu_e)/T}+1}
\int {2 d^3{\vec k}_2\over (2\pi)^3} {1\over e^{(E_2+\mu_e)/T}+1}
{s(E_1+E_2)\over2E_1E_2}\sum_{\vec n} 
\sigma_{e^-e^+\rightarrow \gkk}(s,m_{\vec n}) 
\nonumber\\
&=&{2^{n} I_e(n)\over (n+4)\pi^2} {T^{n+7}\over M_S^{n+2}} \ ,
\label{Qe}
\end{eqnarray}
where $\mu_e$ and $-\mu_e$ are the chemical potentials for electrons and
positrons; $\mu_e\simeq (3\pi^2 n_e)^{1/3}\simeq 345\ {\rm MeV}$
with the electron density $n_e\simeq 1.8\times 10^{38}\ {\rm cm}^{-3}$
at the supernova core. The integral factor is
\begin{equation}
I_e(n)\ =\ \int_0^\infty dx\ dy\ {(xy)^{n/2+2} (x+y)
\over (e^{x-\mu_e/T}+1)(e^{y+\mu_e/T}+1)}\ .
\end{equation}
Numerically, the value of this integral ranges from 0.08 to 86 ($n=2$)
and from 0.62 to 450 ($n=3$) for $T_{\rm SN}$ from 20 MeV to 60 MeV.
We note that the energy-loss rate formula Eq. (\ref{Qe}) 
can also be applied to neutrino-antineutrino annihilation 
at the supernova neutrino sphere.

\subsection{Gravi-Compton-Primakoff scattering}

\begin{figure}[thb]
\epsfysize=1.5in
\epsffile[120 360 400 440]{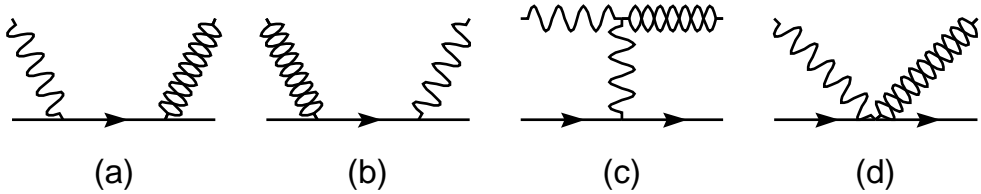}
\caption[]{Feynman diagrams for gravi-Compton-Primakoff scattering.
\label{gv}}
\end{figure}

Figure \ref{gv} represents the Feynman diagrams for the
Gravi-Compton-Primakoff scattering process
$e^-(k_1) + \gamma (q_1) \rightarrow e^-(k_2) + \gkk(q_2)$ 
whose matrix element is
\begin{eqnarray}
i{\cal M}_{\rm GCP} &=& \left({e\kappa\over 2}\right)  
{\overline u}(k_2)\ \biggl[ 
{1\over s-m_e^2} \gamma_\mu k_{2\nu} ({\slashchar k}+m_e) \gamma_\rho +
{1\over u-m_e^2}\gamma_\rho ({\slashchar j}+m_e)\gamma_\mu k_{1\nu}
\nonumber\\
&&+ {2\over t}
(-l\cdot q_1 \gamma_\mu \eta_{\nu\rho} +
\eta_{\mu\rho} \slashchar{q_1} l_\nu 
+ \gamma_\mu l_{\rho} q_{1\nu}
- \gamma_\rho l_\nu q_{1\mu} )- \gamma_\mu \eta_{\nu\rho}
\biggr]\ u(k_1) \epsilon^\rho(q_1) \epsilon^{\mu\nu *}(q_2),
\label{radh}
\end{eqnarray}
where $s=k^2,\ t=l^2,\ u=j^2$ are the Mandelstam variables
and $k=k_1+q_1$, $l=k_1-k_2$, $j=k_1-q_2$;
$\epsilon^\rho(q_1)$ and  $\epsilon^{\mu\nu}(q_2)$ are the polarization
vector and tensor for the photon and KK graviton, respectively.
The first two terms in Eq. (\ref{radh}) 
(Figs. \ref{gv}(a) and \ref{gv}(b)) represent Compton scattering 
and the third term (Fig. \ref{gv}(c))
is the Primakoff process contribution; the last term 
(Fig. \ref{gv}(d)) is due to the contact interaction. 
The Compton and Primakoff processes interfere and can not be separated;
we therefore call this the Gravi-Compton-Primakoff 
(GCP) process.

Since the electron mass is much larger than the temperature of
the Sun and red giant cores where the GCP process is
important, we calculate the cross section in 
the non-relativistic (NR) limit.
We neglect plasma effects in our calculations.

In the NR limit, $T\ll m_e$, we neglect the initial electron
momentum as well as the final-state electron recoil momentum; 
the final KK graviton therefore has the same energy as the incident photon
energy, $\omega$. In the electron rest-frame, the
Mandelstam variables have the following leading order approximation
\begin{equation}
s\ =\ m_e^2 + 2 m_e\omega\ , \qquad
t\ \simeq\ m_{\vec n}^2 - 2 \omega^2 b_{\vec n}\ ,\qquad
u\ \simeq\ m_e^2 - 2 m_e\omega + 2 \omega^2 b_{\vec n}\ ,
\end{equation}
where 
$b_{\vec n}=1-\beta_{\vec n}\cos\theta_{\gamma\vn}$,
$\beta_{\vec n}=\sqrt{1-x_\vn}$,
$x_\vn= m_{\vec n}^2/\omega^2$,
and $\theta_{\gamma\vn}$ is the opening angle between 
the outgoing KK graviton and the incident photon.
The matrix element squared is found to be
(keeping only the leading term in $T/m_e$)
\begin{equation}
 {\sum}'|{\cal M}_{\rm GCP}|^2 
\ =\  {1\over 4} e^2\kappa^2 m_e^2 f_{\rm GCP}
\end{equation}
where
\begin{eqnarray}
f_{\rm GCP} &\simeq& 
{-4\over 3 (x_\vn-2b_\vn)^2 } 
[4 b_\vn^4  - 2 b_\vn^3 (7 + 2 x_\vn) 
+b_\vn^2  (18 + 15 x_\vn + x_\vn^2 )\nonumber\\
&&\qquad\qquad\qquad- 6 b_\vn (2 + 4 x_\vn + x_\vn^2 ) 
+ x_\vn (6 + 9 x_\vn + x_\vn^2 ) ] 
\end{eqnarray}
Neglecting the electron degeneracy,
the volume emissivity is found to be
\begin{equation}
Q_{\rm GCP}\ \simeq\ n_e \int {2d^3{\vec k}\over (2\pi)^3}\ 
{\sum_{\vn}\omega\sigma_{\rm GCP}(\omega,m_\vn)\over e^{\omega/T}-1}
\ \simeq\ {\alpha n_e (4+n)! I_{\rm GCP}(n)\over 2\pi m_e} 
{T^{n+5}\over M_S^{n+2}}
\end{equation}
where $\sigma_{\rm GCP}$ is the cross section for a single KK graviton, and
the summation is over all kinematically accessible KK states with
mass $m_\vn\leq \omega$.
The integral factor is
\begin{equation}
I_{\rm GCP}(n)\ =\ \int_{0}^1 dx_\vn
\ x_\vn^{n/2-1}\int_{1-\beta_\vn}^{1+\beta_\vn} db_\vn\ 
f_{\rm GCP}\ .
\end{equation}
Numerically, the value of this integral is $12.0$ ($6.6$) 
for $n=2$ (3). For the red-giant core, electron
degeneracy is relevant, but we expect this effect
is of order unity, and the limits we derive using the non-degenerate
formula should not be changed significantly.

\subsection{Gravi-bremsstrahlung}

\begin{figure}[thb]
\epsfysize=1.5in
\epsffile[120 360 400 440]{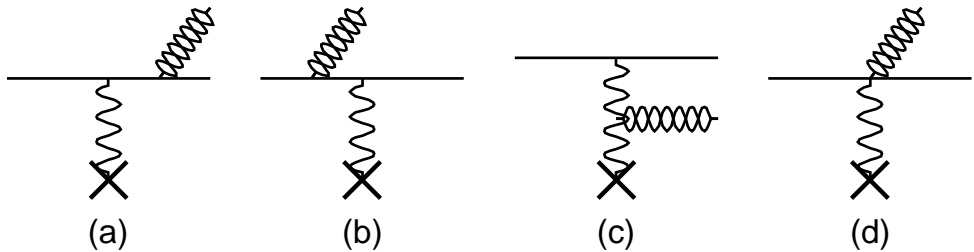}
\caption[]{Feynman diagrams for a bremsstrahlung emission of KK 
gravitons by electrons in the static electric field generated
by nuclei.
\label{gb}}
\end{figure}

Here we consider the bremsstrahlung emission of KK gravitons
by electrons in the static electric field generated by nuclei. 
The diagrams are shown in Fig.~\ref{gb}, 
neglecting those with KK states emitted from the heavy
nuclei, which are suppressed by the large nucleon mass.
The $S$-matrix element for this process is similar to that of 
Eq. (\ref{radh}) and it reads 
\begin{eqnarray}
&&{\cal S}_{\rm GB}\ =\ i 2\pi\delta(k^0_{2}+q_2^0-k_1^0) {\cal M}_{\rm GB},
\nonumber\\ 
&&i {\cal M}_{\rm GB}\ =\ \left({Z e^2\kappa\over 2}\right)  
{\overline u}(k_2)\ \biggl[ 
{1\over s-m_e^2} \gamma_\mu k_{2\nu} ({\slashchar k}+m_e) \gamma_0 +
{1\over u-m_e^2}\gamma_0 ({\slashchar j}+m_e)\gamma_\mu k_{1\nu}+\nonumber\\
&&  {2\over t}
(-l\cdot q_1 \gamma_\mu \eta_{\nu 0} +
\eta_{\mu 0} \slashchar{q_1} l_\nu
+ \gamma_\mu l_0 q_{1\nu}
- \gamma_0 l_\nu q_{1\mu} )- \gamma_\mu \eta_{\nu 0 }
\biggr]\ u(k_1) \biggl[{1\over ({\vec k}_2-{\vec k}_1)^2 + k_S^2}\biggr]
\epsilon^{\mu\nu *}(q_2)\ ,
\label{Sm}
\end{eqnarray}
where $k_S$ in the denominator 
is a screening wave number, corresponding to the electrostatic
potential of a point charge, $e^{-k_S r}/r$.
The delta-function in the $S$-matrix element reflects the conservation
of energy in the static electric field.

We calculate the energy-loss rate in the NR limit, where the initial and
final state electrons have velocities ${\vec\beta}_i$ and ${\vec\beta}_f$,
the  virtual photon has momentum $m_e({\vec\beta}_f-{\vec\beta}_i)$,
and the KK graviton has energy $\omega_\vn={1\over2}m_e(\beta_i^2-\beta_f^2)$.
For the first two terms in Eq. (\ref{Sm}), the leading and 
next-to-leading order terms in the velocity expansion cancel,
so we need to retain all terms in the equation.
The matrix element squared can be factorized as follows
\begin{equation}
{\sum}'|\cM_{\rm GB}|^2\ =\ 
{Z^2 e^4 \kappa^2 f_{\rm GB} 
\over  4 m_e^2 [\bei^2+\bef^2-2\bei\bef c_{if}+k_S^2/m_e^2]^2}\ ,
\end{equation}
where $c_{if}=\cos\theta_{if}$ with $\theta_{if}$ the angle between
the initial and final state electrons; and
\begin{eqnarray}
f_{\rm GB} &\simeq& {-11-26 z^2-11 z^4+48 z c_{if}(1
+z^2 - z c_{if}) \over 3(1-z^2)^2}\nonumber\\
&&+{32 (1+z^2)^2-128 z c_{if} (1+z^2-z c_{if})
\over 3 (1+z^2-2 z c_{if})^2}\ ,
\end{eqnarray}
where $z=\bef/\bei$. The volume emissivity is 
\begin{eqnarray}
Q_{\rm GB} &\simeq& \sum_j {n_e n_j Z_j^2\alpha^2 m_e^{n+1}\bei^{2n+2}
\over 2^{n+1} M_S^{n+2}}
\int_0^1 dx\ x^{n/2-1} \sqrt{1-x} \nonumber\\
&&\times\int_{-1}^1 dc_{if}\int_0^{1}
dz\ z^2 (1-z^2)^{n+2}{y_{\rm GB}\over 
[1+z^2-2 z c_{if}+k_S^2/m_e^2\bei^2]^2}\ ,
\end{eqnarray}
where the summation of $j$ is over all species of nuclei in 
the star. Neglecting $k_S$ and averaging $\bei$ over a
Maxwell-Boltzmann velocity distribution, we have 
\begin{equation}
Q_{\rm GB}\ \simeq\ \sum_j{\Gamma({5\over2}+n) n_e n_j Z_j^2\alpha^2 
I_{\rm GB}(n) 
\over \Gamma({3\over2})} {T^{n+1}\over M_S^{n+2}} \ .
\end{equation} 
The numerical value of the integral 
$I_{\rm GB}(n)$ is 0.7 (0.3) for $n=2\ (3)$.

\subsection{Nucleon-nucleon bremsstrahlung}

\begin{figure}[thb]
\epsfysize=1.5in
\epsffile[88 340 400 432]{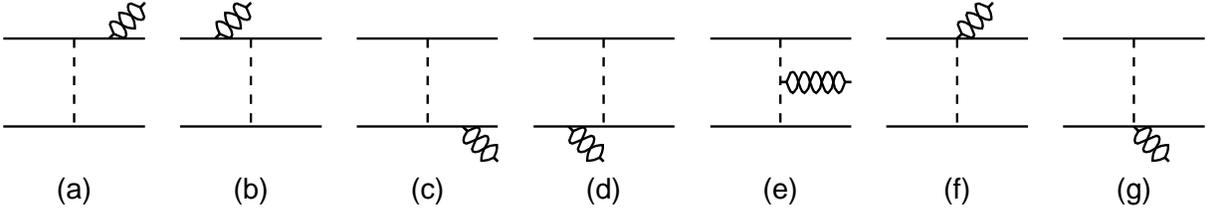}
\caption[]{Representative Feynman diagrams for a nucleon-nucleon 
bremsstrahlung emission of KK gravitons. The contact interaction 
diagrams (f) and (g) are zero since the KK graviton is on-shell.
\label{nb}}
\end{figure}

For the nucleon-nucleon bremsstrahlung process as shown in 
Fig.~\ref{nb}, we take the standard pion-nucleon Yukawa interaction
\begin{equation}
{\cal L}\ =\ -i g_{\pi NN} {\bar N} {\vec\tau}\cdot{\vec\pi}\gamma_5 N\ ,
\end{equation} 
where $g_{\pi NN}\simeq 13.5$ is the pion-nucleon coupling and 
the $\tau_i$ are the Pauli matrices. The contact interaction of a
KK graviton with a nucleon and pion can be derived 
from this Lagrangian. The Feynman rules are
\begin{eqnarray}
&&\pm g_{\pi NN}\gamma_5 \eta_{\mu\nu}\qquad\quad\ \ 
{\rm for}\ nn\pi^0 h^{\vec n}_{\mu\nu}\
{\rm and}\ pp\pi^0 h^{\vec n}_{\mu\nu},\nonumber\\
&&\sqrt{2} g_{\pi NN}\gamma_5 \eta_{\mu\nu}\qquad\quad {\rm for\ the}\ 
np\pi^\pm h^{\vec n}_{\mu\nu}.\nonumber 
\end{eqnarray}
The other relevant Feynman rules can be found in Ref. \cite{hlz}.

It is most convenient to carry out the calculation in the center-of-mass
frame. The initial and final states both have center-of-mass momentum
$\vec{P}$, then all other momenta can be written in terms of
$\vec{P}$ and the relative momenta, ${\vec p}_{1,2}={\vec P}\pm {\vec p}$
and ${\vec p}_{3,4}={\vec P}\pm {\vec q}$.
There are in total 14 diagrams contributing to the nucleon-nucleon
bremsstrahlung process, as shown in Fig. \ref{nb} plus other
7 fermion interchange diagrams.
The four contact interaction diagrams (Fig. \ref{nb}(f,g) and their
interchange diagrams) are automatically zero,
because the KK gravitons are on-shell.
The other 10 diagrams can be grouped into two sets, $A$ and $B$,
with set $A$ the exchange diagrams of set $B$.
The matrix element-squared can be factorized as
(neglecting the pion mass since $m_\pi^2\ll m_N T_{\rm SN}$ in the
supernova core)
\begin{equation}
{\sum}'|{\cM}_{\rm NB}|^2\ =\ 
{1\over4} g_{\pi NN}^4\kappa^2 f_{NN}\ ,
\end{equation}
where in the one-pion exchange approximation
\[ f_{NN} = \left\{ 
\begin{array}{ll}
{1\over 4}\left(|\cM_A|^2 + |\cM_B|^2 - 2|\cM_{A}\cM_{B}|\right) &
\mbox{for $nn$ or $pp$,}\\
                |\cM_A|^2 + 4 |\cM_B|^2 + 4|\cM_{A}\cM_{B}|     &
\mbox{for $np$,}
\end{array}
\right. \]
with (keeping only the leading term in $T_{\rm SN}/m_N$)
\begin{equation}
|\cM_A|^2\ =\ |\cM_B|^2\ \simeq\ 7+9r_{\vec n}\ ,~~~~
|\cM_A\cM_B|\ \simeq\ 4+5r_{\vec n}\ 
\end{equation}
and $r_{\vec n} = m_{\vec n}^2/E_{\vec n}^2$.

The volume emissivity through the nucleon-nucleon bremsstrahlung
process has a relatively simple expression in the non-degenerate 
limit in which the final-state Pauli blocking 
($1-f_{\rm FD}\simeq 1$) is neglected.
Defining  $z_\vn=E_\vn/T$, 
we obtain
\begin{equation}
Q_{\rm NB}\ 
\simeq\ {g_{\pi NN}^4 n_N^2  I_{\rm NB}(n)\over 64\pi^{5/2} m_N^{5/2}}\ 
\ {T^{n+7/2}\over M_S^{n+2}}
\end{equation}
where 
\begin{eqnarray}
I_{\rm NB}(n) &\simeq& \int_0^\infty dz_\vn\ e^{-z_\vn}z_\vn^{n+2} 
(1+{\pi\over4}z_\vn)^{1/2}
\int_0^1 dr_\vn\ r_\vn^{n/2-1} 
(1-r_\vn)^{1/2} f_{NN}\ . 
\end{eqnarray}
Numerically, the value of this integral is about 
80 $(n=2)$ and 280  $(n=3)$ for 
$nn$ and $pp$ bremsstrahlung, and 2700 $(n=2)$ and 9300 $(n=3)$ for 
$np$ bremsstrahlung.

\section{Limits on $M_S$}

We apply our formulae in the previous section to the
energy losses of the Sun, red giants and SN1987A. We briefly 
review the arguments for these cases as follows \cite{PR}: 
\begin{itemize}
\item[(a)] Our Sun is a thermal system with a temperature 
$1.55\times 10^7$ K = 1.3 keV, where thermal pressure 
balances gravity. If the Sun excessively losses energy to the KK 
gravitons, its radius will shrink and the temperature
rise. The Sun would then need to burn more nuclear fuel to compensate for 
the decrease of gravitational energy. This might result in a 
solar age shorter than the current value $4.5\times 10^9$ yr. 
To avoid too rapid consumption of the nuclear fuel, 
a conservative requirement is that the energy-loss
rates from the KK processes do not exceed the solar luminosity,
${\cal L}_\odot=3.90\times 10^{33}\ {\rm erg\ sec}^{-1}$.
\item[(b)] If the core of a red giant near the helium 
flash ($T\sim 8.6$ keV) 
produces excessive KK gravitons, then the helium core may not be
ignited and the star would become a helium white dwarf after 
ascending the red-giant branch, contrary to the observation of
horizontal branch stars. This requires that the KK emission to be
less than the red-giant luminosity at helium flash, 
$\sim 2000\ {\cal L}_\odot$.
\item[(c)] Observational data on SN1987A from IMB and Kamiokande experiments
imply $E\geq 2\times 10^{53}$ ergs emitted over a diffusion 
period of the order of 10 seconds in form of neutrino flux.
This means that much of the binding energy of a neutron star, 
$\sim 3\times 10^{53}$ ergs, is carried away by neutrinos; 
therefore the energy-loss rate from KK states should be less than  
$\sim 10^{52}\ {\rm erg\ sec}^{-1}$.
\end{itemize}
Finally, we note that since the temperatures for the Sun and the 
red-giant core are fairly low, only KK gravitons for the cases 
of $n=2$ and 3 extra dimensions can be efficiently produced there.
In the following we only consider the limits for $n=2$ and 3.

\begin{table}[thb]
\begin{tabular}{|c|c|c|}
(a) & Sun & Red giants \\
 & $n=2\ (\times M^{-4}_{S})$\qquad\qquad\quad  $n=3\ (\times M^{-5}_{S})$ & 
$n=2\ (\times M^{-4}_{S})$\qquad\   $n=3\ (\times M^{-5}_{S})$ \\
\hline
$\dep_\gamma$ & 
$1.7\times 10^{-3}$\qquad\qquad $1.6\times 10^{-11}$ & 
 $6.3$\qquad\qquad  $4.0\times 10^{-7}$ \\
$\dep^{}_{\rm GCP}$ &
$1.3\times 10^{-4}$\qquad\qquad $6.2\times 10^{-13}$ & 
 $50$\qquad\qquad  $1.7\times 10^{-6}$ \\
$\dep^{}_{\rm GB}$  & 
$7.6\times 10^{-4}$\qquad\qquad $1.9\times 10^{-12}$ & 
 $10^3$\qquad\qquad  $1.7\times 10^{-5}$ \\
\end{tabular}
\begin{tabular}{|c|c|c|}
(b) & SN1987A\quad $n=2\ (\times M^{-4}_{S})$ & 
$n=3\ (\times M^{-5}_{S})$ \\
\hline
$\dep_\gamma$ &  $4.7\times 10^{23}\ T_{30}^9$  & 
$1.1\times 10^{20}\ T_{30}^{10}$ \\
$\dep_{e}$ & $8.8\times 10^{17}\quad 1.9\times 10^{21}\quad 
1.9\times 10^{25}$  
&$2.3\times 10^{14}\quad 6.0\times 10^{17}\quad 9.8\times 10^{21}$  \\
$\dep_{\rm NB}$ &  $6.7\times 10^{25}\ T_{30}^{11/2}$  & 
$6.3\times 10^{21}\ T_{30}^{13/2}$ \\ 
\end{tabular}
\vspace{0.2in}
\caption{Energy loss rates (in units of ${\rm erg\ g}^{-1}{\rm sec}^{-1}$)
due to escaping KK gravitons
(a) for the Sun and a red giant from photon-photon annihilation
($\dep_\gamma$), Gravi-Compton-Primakoff scattering
($\dep^{}_{\rm GCP}$) and Gravi-bremsstrahlung ($\dep^{}_{\rm GB}$);
and (b) for a supernova from photon-photon annihilation
($\dep_\gamma$), electron-positron annihilation
($\dep^{}_e$) and nucleon-nucleon bremsstrahlung 
($\dep^{}_{\rm NB}$).
The scaling with $M_S$ (in units of TeV) has been factored out.
The three numbers for $\dep_e$ correspond to the supernova
temperature $T_{\rm SN} = 20, 30, 60$ MeV. 
$T_{30}\equiv T_{\rm SN}/$30 MeV.}
\label{rates}
\end{table}

For the Sun and the red-giant core, we need to consider photon-photon
annihilation, GCP scattering and gravi-bremsstrahlung
processes. The calculated energy loss rates per unit mass
for these three processes are presented in Table~\ref{rates}(a),
scaled with $M_S$ in units of TeV.
We have used the electron densities 
\footnote{All parameters are taken from Ref. \cite{PR}.}
$n_e\simeq6.3\times 10^{25}~{\rm cm}^{-3}$ and 
$3.0\times 10^{29}~{\rm cm}^{-3}$,
and the mass densities 156 g cm$^{-3}$ 
and $10^6$ g cm$^{-3}$ for the Sun and the red-giant core near 
helium flash respectively.
For the case of supernovae, we consider the photon-photon annihilation,
electron-positron annihilation and nucleon-nucleon bremsstrahlung 
processes; those results are shown in Table~\ref{rates}(b).
We take the supernova core density $\simeq 10^{15}$ g cm$^{-3}$
and neutron fraction to be 1.

Using the conservative upper limits on the energy-loss rates of
\begin{equation}
\dep_{\rm Sun}^{}\sim 1\ {\rm erg\ g}^{-1}{\rm sec}^{-1},\quad 
\dep_{\rm RG}^{}\sim 100\ {\rm erg\ g}^{-1}{\rm sec}^{-1}\quad 
{\rm and}\ \  
\dep_{\rm SN}^{}\sim 10^{19}\ {\rm erg\ g}^{-1}{\rm sec}^{-1},
\nonumber
\end{equation}
we obtain the lower limits on $M_S$ summarized in Table \ref{limit}. 

\begin{table}[thb]
\begin{tabular}{|c|c|c|c|}
 $n$  & Sun & Red giant & SN1987A \\
\hline
$2$ & $0.20^{(a)}$,  $0.11^{(c)}$, $0.17^{(d)}$   
& $0.50^{(a)}$,  $0.84^{(c)}$, $1.8^{(d)}$  & 
15 $T_{30}^{2.25(a)}$, $(0.5-37)^{(b)}$, 
$51 T_{30}^{1.375(e)}$,  $(30-130)^{sum}$\\
\hline
$3$ & 
      & 
      &   
1.6 $T_{30}^{2(a)}$, $(0.1-4.0)^{(b)}$, 
$3.6 T_{30}^{1.3~(e)}$,  $(2.1-9.3)^{sum}$ \\
\end{tabular}
\vspace{0.2in}
\caption{Limits to $M_S$ in TeV from (a) photon-photon annihilation,
(b) electron-positron annihilation, (c) gravi-Compton-Primakoff scattering,
(d) gravi-bremsstrahlung and (e) nucleon-nucleon bremsstrahlung. 
The numbers in brackets correspond to the supernova 
temperature range $T_{\rm SN}=20-60$ MeV. For ``sum'',
all contributing processes (a,b,e) are included.
$T_{30}\equiv T_{\rm SN}/30\ {\rm MeV}$.}
\label{limit}
\end{table}

\section{Discussion and conclusion}

We have calculated the energy-loss rates for the Sun, 
red giants and supernovae due to the emission of KK 
gravitons. The lower limits on the string scale $M_S$ can be 
derived by requiring the energy-loss rates to be smaller than 
the respective observed luminosities of those stars. 
We found the lower limits from the Sun and red-giants
are in the range of several hundred GeV with two large extra dimensions. 
The lower limits from the supernova SN1987A are 
more stringent, in particular the nucleon-nucleon 
bremsstrahlung process gives 
\[ M_S \gsim \left\{ 
\begin{array}{ll}
51\ (T_{\rm SN}/30{\rm\ MeV})^{11/8}{\rm\ \ TeV} & \mbox{for $n=2$,}\\
3.6\ (T_{\rm SN}/30\ {\rm MeV})^{13/10}{\rm\ \ TeV} &
\mbox{for $n=3$.}
\end{array}
\right. \]
Our supernova result is consistent with that of 
Ref.~\cite{CP}.

The most important, yet uncertain, parameter in our analyses is
the supernova temperature. We see from Table~\ref{limit} that 
our final results on the $M_S$ limit, including all contributing 
processes in supernovae, range from $30-130$ TeV
($2.1-9.3$ TeV) for $n$=2 (3), corresponding to $T_{\rm SN}=20-60$ MeV.
Although there are other sources of uncertainties in our calculations,
such as in astrophysical parameters, corrections
due to plasma effects, electron degeneracy, screening effects
etc., we do not expect these to significantly alter the $M_S$ limit.
because of the high-power $M_S^{-(n+2)}$ dependence 
of the emission rate. We note that there also exists an interesting 
lower bound from the consideration of the decay of relic graviton 
KK states to photons, which can distort the diffuse photon spectrum \cite{HS}. 
That limit is at the order of $110$ TeV for $n=2$, but it has
uncertainties associated with cosmological models.

The lower limits obtained from the astrophysical processes 
can be complementary to those from collider experiments 
\cite{hlz,nima,collider}, in particular to the collider processes
with virtual KK state exchange, which has the string scale 
dependence of $M_S^{-4}$, essentially independent of $n$.
On the other hand, a string scale at the order of 50 TeV 
for $n=2$ would make string effects inaccessible at collider experiments. 
From Eq.~(\ref{rel}), this scale corresponds to two compact dimensions 
of the size of about $10^{-4}$ mm, which is beyond the sensitivity of the
tabletop gravitation experiments being planned \cite{gravi}. 
Finally, from theoretical point of view, such a high scale 
for two extra dimensions makes the KK scenario 
less attractive since one of the motivations for introducing a
TeV-scale string theory is to solve the weak-GUT scale hierarchy problem.

\vskip 0.2in
{\it Acknowledgments}: 
We would like to thank Wayne Repko for a comment on our manuscript. 
This work was supported in part by a DOE grant
No. DE-FG02-95ER40896 and in part by the Wisconsin Alumni Research 
Foundation. 

\bibliographystyle{unsrt}

\end{document}